\documentclass[aps, prb, superscriptaddress,preprintnumbers,amssymb, preprint]{revtex4-2}
\usepackage{graphicx}
\usepackage{dcolumn}
\usepackage{bm}

\begin{document}

\preprint{ver 8}

\title{Anomalous High-Field Magnetotransport in CaFeAsF due to the Quantum Hall Effect}



\author{Taichi Terashima}
\email{TERASHIMA.Taichi@nims.go.jp}
\affiliation{International Center for Materials Nanoarchitectonics, National Institute for Materials Science, Tsukuba 305-0003, Japan}
\author{Hishiro T. Hirose}
\affiliation{Research Center for Functional Materials, National Institute for Materials Science, Tsukuba 305-0003, Japan}
\author{Naoki Kikugawa}
\affiliation{Center for Green Research on Energy and Environmental Materials, National Institute for Materials Science, Tsukuba, Ibaraki 305-0003, Japan}
\author{Shinya Uji}
\affiliation{International Center for Materials Nanoarchitectonics, National Institute for Materials Science, Tsukuba 305-0003, Japan}
\author{David Graf}
\affiliation{National High Magnetic Field Laboratory, Florida State University, Tallahassee, FL 32310, USA}
\author{Takao Morinari}
\affiliation{Course of Studies on Materials Science, Graduate School of Human and Environmental Studies, Kyoto University, Kyoto 606-8501, Japan}
\author{Teng Wang}
\author{Gang Mu}
\email{mugang@mail.sim.ac.cn}
\affiliation{State Key Laboratory of Functional Materials for Informatics, Shanghai Institute of Microsystem and Information Technology, Chinese Academy of Sciences, Shanghai 200050, China}
\affiliation{CAS Center for Excellence in Superconducting Electronics (CENSE), Shanghai 200050, China}


\date{\today}



\maketitle

\newpage

\section*{Abstract}
CaFeAsF is an iron-based superconductor parent compound whose Fermi surface is quasi-two dimensional, composed of Dirac-electron and Schr\"odinger-hole cylinders elongated along the $c$ axis.
We measured the longitudinal and Hall resistivities in CaFeAsF with the electrical current in the $ab$ plane in magnetic fields up to 45 T applied along the $c$ axis and obtained the corresponding conductivities via tensor inversion.
We found that both the longitudinal and Hall conductivities approached zero above $\sim$40 T as the temperature was lowered to 0.4 K.
Our analysis indicates that the Landau-level filling factor is $\nu$ = 2 for both electrons and holes at these high field strengths, resulting in a total filling factor $\nu$ = $\nu_{\mathrm{hole}} - \nu_{\mathrm{electron}}$ = 0.
We therefore argue that the $\nu$ = 0 quantum Hall state emerges under these conditions.

\section*{Introduction}

Since the discovery of iron-based superconductors \cite{Kamihara08JACS}, researchers have attempted to elucidate the electronic structure of the antiferromagnetic parent phase from which the high-$T_{\mathrm{c}}$ superconductivity emerges.
The antiferromagnetic parent phase is reached via a structural transition from tetragonal to orthorhombic and an antiferromagnetic transition \cite{Cruz08nature, Rotter08PRB}.
Superconductivity emerges when the antiferromagnetic parent phase is suppressed by chemical substitution or pressure application \cite{Kamihara08JACS, Rotter08PRL,Sasmal08PRL, Johnston10AdvPhys}.
It was soon recognized that owing to a peculiar topological feature of the electronic band structure, the spin-density-wave gap in the antiferromagnetic parent phase cannot be fully open; instead, it should have a node that gives rise to Dirac fermions \cite{Ran09PRB, Morinari10PRL}.

CaFeAsF is a variant of 1111 iron-based superconductor parent compounds, in which the LaO in LaFeAsO is replaced with CaF \cite{Matsuishi08JACS}.
Unlike LaFeAsO, large high-quality single crystals of CaFeAsF can be grown \cite{Ma15SST}; hence, it is suitable for studying electronic structure.
Our previous study demonstrates that the Fermi surface of CaFeAsF comprises a symmetry-related pair of $\alpha$ Dirac electron cylinders occurring at the gap nodes and a $\beta$ Schr\"odinger hole cylinder at the Brillouin zone center (Fig.~1\textbf{c}) \cite{Terashima18PRX}.
The carrier density is as small as $\sim$10$^{19}$ cm$^{-3}$, which corresponds to 10$^{-3}$ per Fe (Table 1).
We note that it is of comparable magnitude to carrier densities reported for EuMnBi$_2$ \cite{Masuda16SciAdv} and highly doped Bi$_2$Se$_3$ \cite{Cao12PRL}, both of which exhibit bulk quantum Hall effect.
As expected from the Fermi surface, CaFeAsF exhibits highly two-dimensional electronic conduction with an anisotropy ratio of 200--300 between the $c$-axis and $ab$-plane resistivities, although the temperature dependence is metallic for both directions in the antiferromagnetic phase.
Previously, Ma \textit{et al.} performed resistivity measurements on CaFeAsF up to $B$ = 65 T \cite{Ma18scichina}.
They found that the resistivity increased sharply for fields above $B \approx 30$ T applied along the $c$ axis at low temperatures and that the temperature dependence became nonmetallic in the high-field region.
Based on a scaling analysis, they argued that a metal--insulator quantum phase transition occurred at $B_{\mathrm{c}}$ = 30 T.

In the present work, we measured not only the longitudinal resistivity $\rho_{xx}$ but also the Hall resistivity $\rho_{yx}$ in CaFeAsF up to $B$ = 45 T.
We determined the longitudinal $\sigma_{xx}$ and Hall $\sigma_{xy}$ conductivities by tensor inversion and found that both approached zero above about 40 T.
The temperature dependence of the longitudinal resistivity $\rho_{xx}$ at those high fields is rather weak, approximately $T^{-1}$, and shows a tendency of saturation as $T \rightarrow 0$, which is at variance with energy-gap formation due to e.g., a spin- or charge-density wave or the magnetic freeze-out. 
We argue that the anomalous high-field insulating phase is the $\nu$ = 0 quantum Hall state.

\section*{Results}
\subsection*{Shubnikov--de Haas oscillation}

Single crystals of CaFeAsF were prepared at the SIMIT in Shanghai via a CaAs self-flux method \cite{Ma15SST}.
We first measured Shubnikov--de Haas (SdH) oscillations in the main sample (named D2) used for the present study, employing a 20-T superconducting magnet and a dilution refrigerator at the NIMS in Tsukuba to determine the Fermi-surface parameters (Fig.~1).
Two frequency peaks corresponding to the $\alpha$ Dirac electron and $\beta$ Schr\"odinger hole cylinders appear in the Fourier transform of the second-derivative curve as a function of 1/$B$ (Fig.~1\textbf{b}).
The fact that neither of the frequencies splits into a doublet confirms the highly two-dimensional electronic structure.
We determined the effective masses $m^*$ for $B \parallel c$ from the temperature dependences of the peak amplitudes.
To extract the frequencies $F$ and Dingle temperatures $T_{\mathrm{D}}$, we then fitted the Lifshitz--Kosevich formula \cite{Richards73PRB, Shoenberg84} to the lowest-temperature data with the effective masses fixed (see \cite{Terashima22npjQM} for details of the fitting procedure).
The obtained parameters (Table 1) are consistent with our previous reports \cite{Terashima18PRX, Terashima22npjQM} (the sample of \cite{Terashima18PRX} is the same as sample X1 used below).
The spin-splitting parameter $S = (1/2)g(m^*(\theta)/m_{\mathrm{e}})$ \cite{Shoenberg84} is based on \cite{Terashima18PRX}, which showed from the field-angle dependence of the SdH oscillations that $S$ = 1/2 and 5/2 for $\alpha$ and $\beta$, respectively, at $\theta \approx 50^{\circ}$.
Here, $\theta$ is the field angle measured from the $c$ axis.
We obtain the tabulated values $S$ for $\theta = 0$ assuming the relation $m^*(\theta) = m^*(0)/\cos \theta$, which holds for cylindrical Fermi surfaces.

We prepared a Landau-level diagram based on these parameters, neglecting a possible small $c$-axis dispersion (Fig.~1\textbf{d}).
The shading around the levels represents the Landau-level broadening $\Gamma$ due to impurity scattering derived from the Dingle temperature.
While the shaded areas overlap with each other under low fields, the Landau levels are well separated near and above $B$ = 40 T.
We note that, although Fig 1\textbf{d} is based on the Fermi-surface parameters of sample D2 (Table I), the parameters of sample X1 \cite{Terashima18PRX} give essentially the same Landau-level diagram, especially near the Fermi level.

We determined the Fermi level by putting electrons in Landau levels from the lowest-energy one until the carrier conservation, i.e., $n_{\mathrm{e}}(B) - n_{\mathrm{h}}(B) = n_{\mathrm{e}}(0) - n_{\mathrm{h}}(0)$, is reached. Here $n_{\mathrm{e(h)}}(B)$ is the electron (hole) density at a magnetic-field strength $B$, and $n_{\mathrm{e}}(0) - n_{\mathrm{h}}(0) = -0.35\times 10^{-3}$ per Fe  from Table 1.
The determined Fermi level is plotted above $B$ = 20 T (dashed line).
Since we neglected the Landau-level broadening and assumed perfect two dimensionality, the Fermi level coincides with a Landau level.
If we take the broadening into account, the Fermi level traverses the width of a broadened Landau level from the low-energy side to the high-energy one as the magnetic field is varied and hence crosses the center of the level at a certain field:
for example, the Fermi level crosses the center of the $\beta$0$+$ level from below to above at $B$ = 38 T.
This corresponds to that the $\beta$0$+$ level becomes less-than-half filled above $B$ = 38 T in terms of the hole carrier.

\subsection*{Longitudinal and Hall conductivities}

Figure~2 shows our main result.
We simultaneously measured the longitudinal resistivity $\rho_{xx}$ and Hall resistivity $\rho_{yx}$ for sample D2 in fields up to $B$ = 45 T applied along the $c$ axis using a hybrid magnet at the NHMFL in Tallahassee.
To eliminate mixing of the two components, we performed measurements in both positive and negative fields by rotating the sample by 180$^{\circ}$ and symmetrized and antisymmetrized $\rho_{xx}$ and $\rho_{yx}$, respectively.
As reported previously \cite{Ma18scichina}, the longitudinal resistivity increases rapidly above $B \approx 30$ T as the temperature is reduced.
The value of $\rho_{xx}$ at $B$ = 45 T and $T$ = 0.4 K is $2.7 \times 10^2$ times larger than the zero-field value.
Because an anomalous resistivity increase is also observed for $\rho_{zz}$ (i.e., $I \parallel B \parallel c$), as measured on sample X2 and shown in Fig.~4\textbf{b}, it cannot be ascribed to classical magnetoresistance arising from the orbital motions of the carriers.

The Hall resistivity above $B \approx 30$ T is positive at $T$ = 10 and 5 K but becomes negative at $T$ = 1.6 and 0.4 K (Fig.~2\textbf{b}).
Within the classical magnetotransport theory, the Hall resistivity in the high-field limit is determined by the net number of carriers \cite{Hurd72}:
$\rho_{yx} = B/(e(p-n))$, where $n$ and $p$ are the densities of electrons and holes, respectively.
This gives $\rho_{yx}$ = 5.2 m$\Omega$cm at $B$ = 45 T.
Thus, the observed sign change is difficult to explain within the semiclassical theory.
In addition, the magnitude of the experimental Hall resistivity is considerably smaller than the theoretical expectation at all temperatures.
In summary, neither $\rho_{xx}$ nor $\rho_{yx}$ above $B \approx 30$ T can be explained by the classical theory.

Figure~2\textbf{c} shows the longitudinal conductivity $\sigma_{xx}$ and the Hall conductivity $\sigma_{xy}$, which are given respectively by $\sigma_{xx} = \rho_{xx}/(\rho_{xx}^2 + \rho_{yx}^2)$ and $\sigma_{xy} = \rho_{yx}/(\rho_{xx}^2 + \rho_{yx}^2)$.
In this figure, we also show the conductivity corresponding to one conductance quantum $e^2/h$ per FeAs layer. 
This figure gives a better perspective on the high-field magnetotransport in CaFeAsF.
Note that both $\sigma_{xx}$ and $\sigma_{xy}$ at $B$ = 45 T are much smaller than $e^2/h$ per FeAs layer, and they approach zero as the temperature is reduced.

\subsection*{Landau-level-crossing fields}

Before discussing the anomalous high-field state, let us confirm our Landau-level diagram (Fig.~1\textbf{d}) by analyzing the longitudinal conductivity $\sigma_{xx}$.
Figure~3 shows $\sigma_{xx}$ and its second field derivative as a function of 1/$B$.
The Landau-level-crossing fields for $\alpha$ and $\beta$ based on Fig.~1\textbf{d} are indicated by green and red triangles, respectively.
The hypothetical spinless crossing fields (i.e, without Zeeman energy) are also indicated by circles.
For $1/B > 0.05$ T$^{-1}$ ($B < 20$ T), the local minima of $\mathrm{d}^2 \sigma_{xx}/\mathrm{d}B^2$, which correspond to the local maxima of $\sigma_{xx}$, coincide with the spinless crossing fields for $\beta$ (red circles).
The $(n-1)$-th up-spin crossing field (upward triangle) and $(n+1)$-th down-spin one (downward triangle) are close to each other, and the $n$-th spinless crossing field is halfway between them.
This explains the above observation and also the absence of clear spin-splitting of oscillation peaks.
The existence of the $\alpha$ oscillation is manifested in the fact that the maximum of $\mathrm{d}^2 \sigma_{xx}/\mathrm{d}B^2$ at $1/B$ = 0.105 (or the minimum at $1/B$ = 0.158) is lower than the neighboring ones.
The splitting of up-spin and down-spin crossing fields is small compared to the oscillation period, which explains the absence of clear spin-splitting of oscillation.
For $1/B < 0.05$ T$^{-1}$ ($B > 20$ T), the behavior of $\mathrm{d}^2 \sigma_{xx}/\mathrm{d}B^2$ is not directly related to the crossing fields likely because the Fermi energy is no more constant as indicated in Fig.~1\textbf{d}.

\section*{Discussion}

We can now discuss the anomalous high-field state based on the Landau-level diagram Fig.~1\textbf{d}.
For the $\alpha$ Dirac electrons, only the lowest Landau levels, $\alpha 0+$ and $\alpha 0-$, are occupied above $B \approx 30$ T.
The filling factor $\nu_{\alpha}$ there is calculated to be 2;
the lowest Landau level of the Dirac electrons is special and contributes 1/2 to the filling factor.
Because there are two $\alpha$ cylinders in the Brillouin zone and two spin species, the sum of their contributions becomes 2.
For the $\beta$ holes, the $\beta$0+ level becomes less-than-half filled above $B \approx 38$ T, and hence, the Fermi level exists among localized tail states.
Accordingly, the two fully filled levels $\beta$0- and $\beta$1- give the filling factor $\nu_{\beta}$ = 2.
As the total filling factor $\nu =  \nu_{\beta}-\nu_{\alpha}$ \cite{Mendez85PRL} becomes zero above $B \approx 38$ T, we expect both $\sigma_{xx}$ and $\sigma_{xy}$ to approach zero as the quantum Hall state develops.
This explains our observations.

For the quantum Hall effect to operate, Landau levels need to be separated well from each other.
Figure 1\textbf{d} shows that this prerequisite is satisfied above $B \approx 38$ T:
the shaded areas around the levels, which represent broadening of the levels, do not overlap.
On the other hand, the shaded areas overlap at lower fields.
For example, the shaded areas around $\beta$0$+$ and $\beta$2$-$ overlap until $B \approx 31$ T.
This prevents the quantum Hall effect to show up at low fields.
Accordingly, $\rho_{yx}$ exhibits usual SdH oscillations rather than quantum Hall plateaus at low fields (Fig. 2\textbf{b}).

Usually, in quantum Hall states ($\nu \neq 0$), $\sigma_{xx}$ = 0 but $\sigma_{xy}$ remains finite (= $\nu e^2/h$).
Therefore, $\rho_{xx}$ = 0 and $\rho_{yx} = h /(\nu e^2)$.
However, when $\nu = 0$, $\sigma_{xy}$ is also expected to approach zero.
Accordingly, the way $\rho_{xx}$ and $\rho_{yx}$ behave depends on the detailed behavior of $\sigma_{xx}$ and $\sigma_{xy}$.
The $\nu = 0$ quantum Hall state has been studied in InAs/GaSb systems \cite{Daly96PRB, Nicholas00PRL}, HgTe quantum wells \cite{Gusev10PRL}, and graphene devices \cite{Abanin07PRL, Checkelsky08PRL}.
In the first two systems, electrons and holes coexist, while in graphene the $\nu$ = 0 state is achieved by tuning the Fermi level to the Dirac point.
The reported observations may be summarized as follows:
In the InAs/GaSb systems, where imbalances exist between the electron and hole carriers, $\sigma_{xy}$ is much closer to zero than $\sigma_{xx}$ at $\nu$ = 0. 
In the latter two systems, $\sigma_{xy}$ even crosses zero as the (net) carrier polarity is varied by the gate voltage.
However, although $\rho_{xx}$ at $\nu$ = 0 increases strongly with the magnetic field as the temperature is reduced, the temperature dependence is much weaker than that due to thermal activation and exhibits saturation at low temperatures, suggesting that $\sigma_{xx}$ remains finite as $T \to 0$.


We show the temperature dependence of $\rho_{xx}$ at $B$ = 45 T as a function of temperature in Fig.~4\textbf{a}.
Because sample D2 was measured only at four temperatures (the blue circles), we also plotted data from sample X1 (green squares).
In the latter case, the magnetic field was applied at 9.2$^{\circ}$ off the $c$ axis (for technical reasons), but this level of field tilting affects the resistivity only slightly, as can be seen in Fig.~4\textbf{b}. 
The temperature dependence cannot be described by an activation form (inset); rather, it follows a $T^{-1}$ dependence over a wide temperature range and then saturates at the lowest temperatures.
(Ma \textit{et al.} observed a stronger temperature dependence up to $T^{-1.7}$ at higher fields, $B \approx 60 $ T \cite{Ma18scichina}.)
This is similar to the behavior observed in the InAs/GaSb systems and in HgTe quantum wells \cite{Daly96PRB, Gusev10PRL}.
In those studies, the origin of the $T^{-1}$ behavior was discussed in connection with possible roles played by disorder but is not yet settled.
We also examined the temperature dependence of the $c$-axis resistivity $\rho_{zz}$ measured in sample X2 (Fig. 4\textbf{a}, red circles).
Although the $c$-axis conduction mechanism is unclear presently, $\rho_{zz}$ exhibits essentially the same temperature dependence.

We also note additional similarities between the InAs/GaSb systems and CaFeAsF.
The Hall resistivity in the InAs/GaSb systems shows stronger SdH oscillations than the longitudinal resistivity.
It shows neither a plateau nor a plateau-to-plateau transition, except for one Hall plateau at $\nu$ = 1 \cite{Nicholas00PRL}.
The sign of the Hall resistivity near the high-field region $\nu$ = 0 changes suddenly from positive to negative as the temperature is lowered from 1.8 to 0.5 K (the lower inset of Fig. 1 of \cite{Nicholas00PRL}).
For comparison, the Hall resistivity in CaFeAsF shows clear oscillations but neither a plateau nor a plateau-to-plateau transition (Fig. 2\textbf{b}).
Above $B \approx 30$ T, the sign of the Hall resistivity changes as the temperature is reduced from 5 to 1.6 K.

We now contrast the present observations with a magnetic-field-induced metal-insulator transition in doped semiconductors \cite{Shklovskii84Book}, most typically observed in $n$-type InSb \cite{Gershenzon74SovPhysSemiconduct, Ishida77JPSJ, Mansfield78JPC, Tokumoto80SSC}.
The metal in the latter case means that the resistivity at zero field does not diverge as $T \to 0$, being roughly constant at low temperatures, although it is of an activation type at high temperatures.
As the magnetic field is applied to such a `metallic' state, the wave functions of impurity electrons are squeezed and more and more localized at donor sites.
This leads to a rapid decrease in the overlap of neighboring wave functions and hence to an exponential increase in the resistivity.
The longitudinal resistivity often increases by several orders of magnitude or more, and the Hall resistivity is also enhanced \cite{Ishida77JPSJ, Mansfield78JPC}.
This process is called magnetic freeze-out.
At sufficiently low temperatures, the temperature dependence of the resistivity in magnetic fields is described by a general variable-range hopping formula $\rho = \rho_0 \exp[(T_0/T)^p]$, where $0 < p < 1$ \cite{Tokumoto80SSC, Biskupski91JPCM}.
Theoretically, $p$ = 1/4 and 1/3 for Mott's law in three- and two-dimensions and 1/2 with the so-called Coulomb gap at $E_{\mathrm{F}}$ \cite{Shklovskii84Book}.
On the other hand, CaFeAsF is a semimetal:
carriers are intrinsic, and there is no donor site where carriers can be localized.
Although the longitudinal resistivity increases more than two orders of magnitude at $B$ = 45 T, the magnitude of the Hall resistivity remains smaller than that expected from the carrier concentration.
The temperature dependence of the longitudinal resistivity is weak, $\sim T^{-1}$, and saturates at lowest temperatures (Fig.~4\textbf{a}).
It cannot be described by an activation form nor a variable range hopping with $p$ = 1/4, 1/3, or 1/2.
Thus, the present observations cannot be explained by the magnetic freeze-out.

Finally, we mention the field-angle dependence of the resistivity increase.
Figure~4\textbf{b} shows the longitudinal resistivity in sample X1 for various field directions as a function of the $c$-axis component of the applied field $B\cos \theta$.
The curves do not collapse into a single curve, as noted by \cite{Ma18scichina}.
This is essentially because the Zeeman energy is not determined by $B \cos \theta$, but by $B$.

There seems to be an interplay between the SdH oscillations and the $\nu$ = 0 quantum Hall state.
The SdH oscillations from up-spin and down-spin electrons interfere, resulting in the spin reduction factor $R_{\mathrm{s}}=\cos (\pi S)$ in the oscillation function \cite{Shoenberg84}.
Because $S$ increases with $\theta$ and becomes 1/2 and 5/2 for $\alpha$ and $\beta$, respectively, at $\theta \approx 50^{\circ}$, $R_{\mathrm{s}}$ changes sign there for both oscillations.
This can be confirmed in the inset of Fig.~4\textbf{b}:
the SdH-oscillatory resistivity exhibits a minimum at $B\cos \theta \approx 19$ T for $\theta < 50^{\circ}$, while a maximum for $\theta > 50^{\circ}$.
(This is a combined effect of the sign changes of $\alpha$ and $\beta$.
If we like to confirm a sign change for $\alpha$ and $\beta$ separately, we have to filter the signals to separate the two frequencies, which was performed in \cite{Terashima18PRX}.)
Although the SdH-oscillatory resistivity in the $B\cos \theta$ = 40--45 T region is positive at small $\theta$, as $\theta$ increases, it eventually becomes negative because of the sign change of $R_{\mathrm{s}}$.
Thus the resistivity increase above $B \cos \theta \approx 30$ T is suppressed as $\theta$ is increased, and larger magnetic fields are needed to reach the $\nu$ = 0 quantum Hall state.

In summary, we measured both longitudinal and Hall resistivities in CaFeAsF up to $B$ = 45 T.
By tensor inversion, we obtained $\sigma_{xx}$ and $\sigma_{xy}$ and found that both were suppressed to nearly zero above $B \approx 40$ T.
The Landau-level diagram constructed from the analysis of the SdH oscillations indicates that the filling factor is two for both electrons and holes at those high fields and hence that the total is zero.
We therefore argued that a $\nu$ = 0 quantum Hall state appeared at those fields.
The quantum Hall effect in bulk materials has been reported, for example, for organic Bechgaard salts \cite{Chaikin83PRL83, Ribault84JPL, Cooper89PRL, Hannahs89PRL}, $\eta$-Mo$_4$O$_{11}$ \cite{Hill98PRB}, EuMnBi$_2$ \cite{Masuda16SciAdv}, highly doped Bi$_2$Se$_3$ \cite{Cao12PRL}, and ZrTe$_5$ \cite{Tang19Nature}.
Interestingly, in the Bechgaard salts spin-density-wave formation removes most of the Fermi surface, leaving only tiny two-dimensional pockets, which give rise to the quantum Hall effect.
This is analogous to the case of CaFeAsF, in which the nodal spin-density-wave leaves tiny cylindrical Fermi pockets, which exhibit the quantum Hall effect.
However, the $\nu$ = 0 state has been studied only in two-dimensional systems thus far.
CaFeAsF is the first bulk material to exhibit the $\nu$ = 0 state.
Studies comparing CaFeAsF with two-dimensional systems are thus expected to be fruitful.

\section*{Methods}
\subsection*{Samples and measurements}
CaFeAsF single crystals D2 (1 x 0.7 x 0.05 mm$^3$), X1 (1.7 x 0.6 x 0.05 mm$^3$), and X2 (0.7 x 0.3 x 0.1 mm$^3$) were prepared by a CaAs self-flux method at the SIMIT in Shanghai as described in \cite{Ma15SST}.
The electrical contacts were spot-welded and reinforced with silver conducting paint.
For magnetotransport measurements, we used a 20-T superconducting magnet and a dilution refrigerator with a base temperature of 0.03 K at the NIMS in Tsukuba, and the 45-T hybrid magnet and a $^3$He insert at the NHMFL in Tallahassee.
To eliminate mixing of $\rho_{xx}$ and $\rho_{yx}$, measurements were performed in both positive and negative fields and experimental signals $\rho_{xx}^{\mathrm{exp}}$ and $\rho_{yx}^{\mathrm{exp}}$ were symmetrized and antisymmetrized to obtain true $\rho_{xx}$ and $\rho_{yx}$, respectively:
i.e., $\rho_{xx}(B) = (\rho_{xx}^{\mathrm{exp}}(B) + \rho_{xx}^{\mathrm{exp}}(-B))/2$, and $\rho_{yx}(B) = (\rho_{yx}^{\mathrm{exp}}(B) - \rho_{yx}^{\mathrm{exp}}(-B))/2$.
(In the case of the hybrid magnet, the magnetic-field direction cannot be reversed, and hence the sample was rotated by 180$^{\circ}$.)

\section*{Data availability}
The data that support the findings of this study are available from the corresponding authors upon reasonable request.

\section*{Competing interests}
The authors declare no Competing Interests.

\section*{Author contributions}
TT planned the project and wrote the manuscript.  TW and GM prepared the samples.  TT, HTH, NK, SU, and DG performed the measurements.  TT and TM analyzed the data.

\section*{acknowledgments}
This work was supported in Japan by JSPS KAKENHI Grant Numbers 17K05556, 22H04485 and 22K03537.
This work was supported in China by the Youth Innovation Promotion Association of the Chinese Academy of Sciences (No. 2015187).
A portion of this work was performed at the National High Magnetic Field Laboratory, which is supported by the National Science Foundation Cooperative Agreement No. DMR-1644779 and the State of Florida.

\section*{References}

\newpage

\begin{figure*}
\includegraphics[width=10cm]{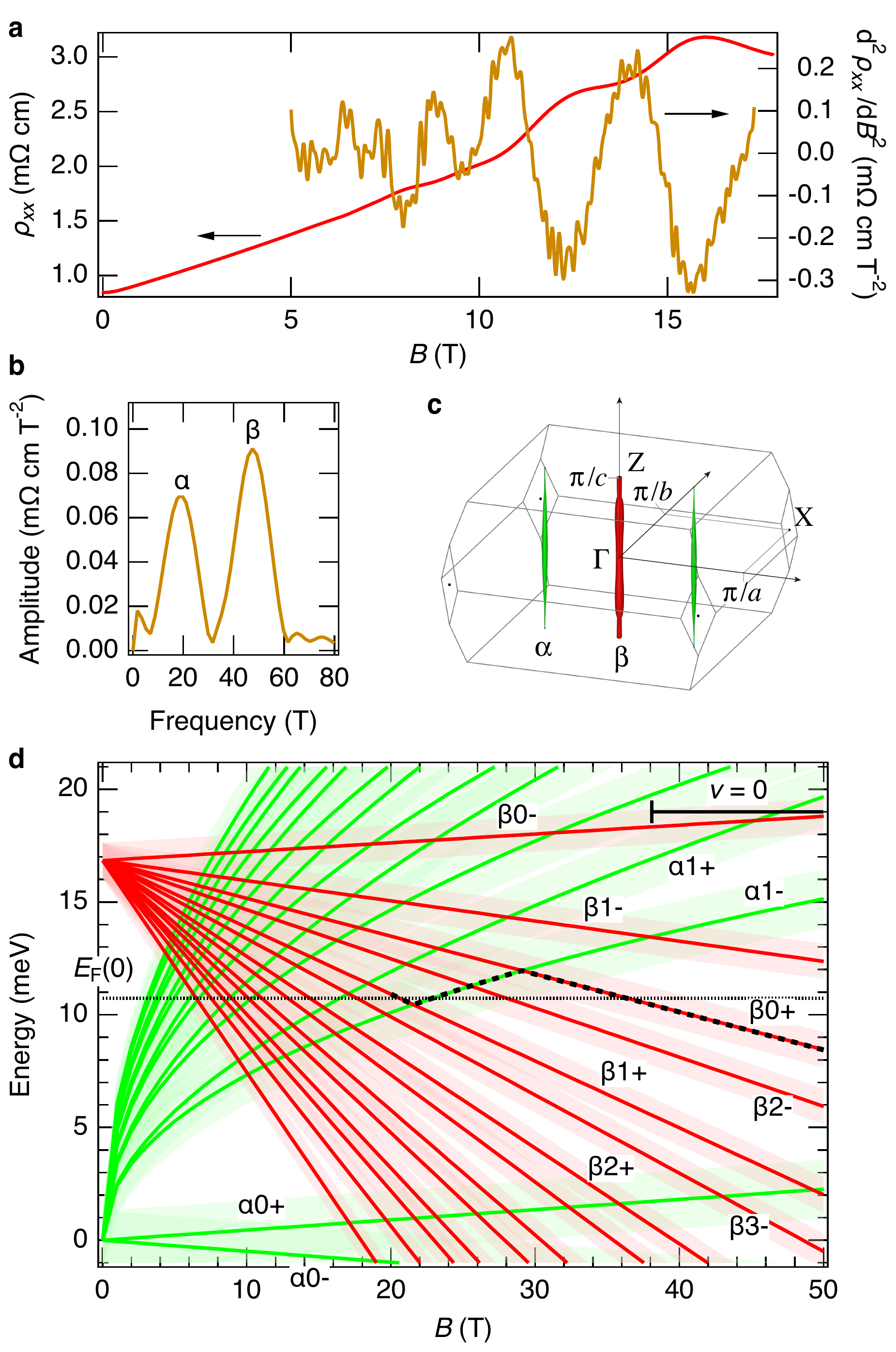}
\caption{\label{Fig1}
\textbf{Shubnikov-de Haas oscillation, Fermi surface, and Landau-level diagram in CaFeAsF.}
\textbf{a} In-plane resistivity $\rho_{xx}$ as a function of the magnetic field $B$ applied along the $c$ axis, as measured on sample D2 at $T$ = 0.03 K. The corresponding second-derivative curve, $\mathrm{d}^2 \rho_{xx} / \mathrm{d} B^2$, is also shown.  \textbf{b} Fourier transform of the curve $\mathrm{d}^2 \rho_{xx} / \mathrm{d} B^2$ vs $1/B$. The labels $\alpha$ and $\beta$ refer to electrons and holes, respectively.  \textbf{c} Fermi surface of CaFeAsF \cite{Terashima18PRX}.  \textbf{d} Landau-level diagram for CaFeAsF calculated from the Fermi-surface parameters in Table 1. The shading around the levels is based on the Landau-level broadening $\Gamma$. The black dashed line above $B$ = 20 T shows the field-dependent Fermi level expected from the carrier conservation when level broadening is neglected (see text). Some of the levels are labeled with the carrier type ($\alpha$ or $\beta$), the Landau-level index (0, 1, etc.), and the spin species (+ or -). The horizontal bar shown at the upper right indicates the high-field region where $\nu$ = 0.
}
\end{figure*}

\begin{figure*}
\includegraphics[width=10cm]{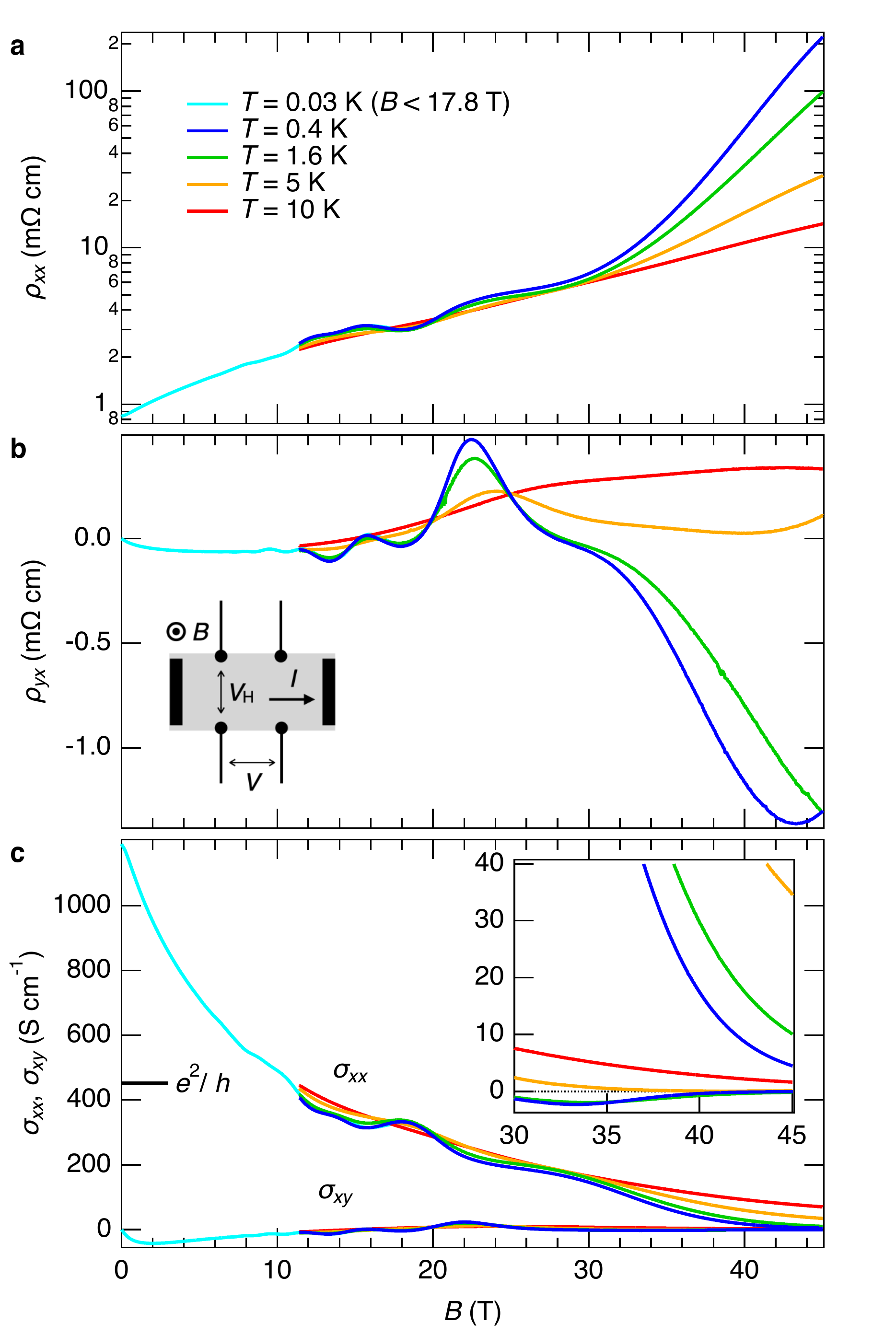}
\caption{\label{Fig2}
\textbf{Magnetotransport in CaFeAsF in fields up to $B$ = 45 T applied along the $c$ axis.} 
\textbf{a} The longitudinal resistivity $\rho_{xx}$ and \textbf{b} the Hall resistivity $\rho_{yx}$. The inset in \textbf{b} shows a schematic of the sample.  \textbf{c} The longitudinal conductivity $\sigma_{xx}$ and Hall conductivity $\sigma_{xy}$. The conductivity corresponding to one conductance quantum $e^2/h$ per FeAs layer is indicated. The inset shows the blow-up of the high-field region.  In \textbf{a}, \textbf{b}, and \textbf{c}, the $T$ = 0.03 K curves (cyan) almost coincide with the $T$ = 0.4 K ones (blue) and hence almost indiscernible above $B$ = 11.4 T.
}
\end{figure*}

\begin{figure*}
\includegraphics[width=10cm]{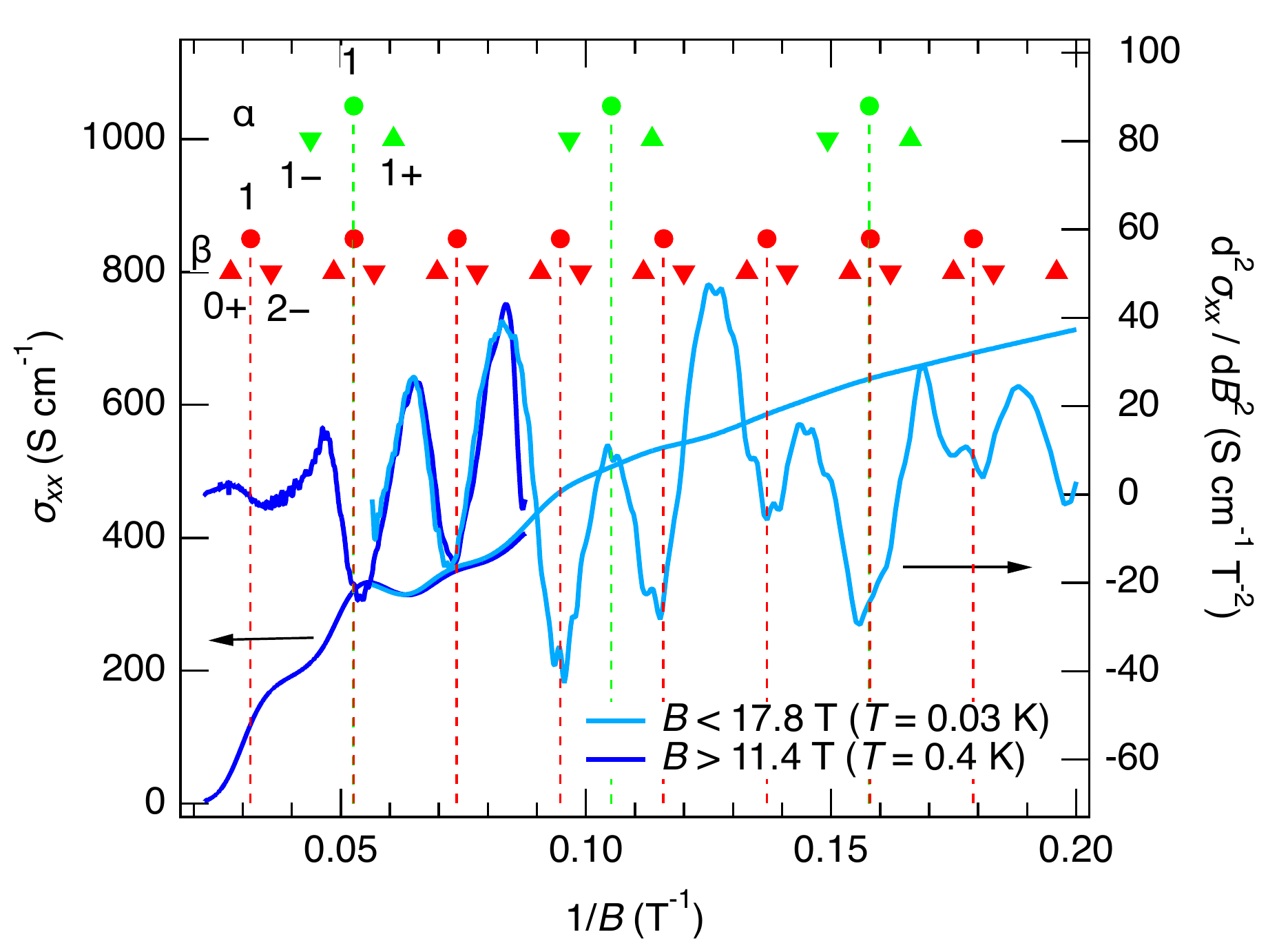}
\caption{\label{Fig3}
\textbf{Shubnikov-de Haas oscillations and Landau-level crossing fields.}  
The longitudinal conductivity in CaFeAsF and its second field derivative are shown as a function of $1/B$.  The green (red) triangles indicate the Landau-level crossing fields for $\alpha$ ($\beta$) based on Fig.~1\textbf{d}.   The hypothetical spinless crossing fields (i.e, without Zeeman energy) are also indicated by circles.  The labels attached to marks are the Landau index (0, 1, etc.) and spin species (+/-- or none).
}
\end{figure*}

\begin{figure*}
\includegraphics[width=10cm]{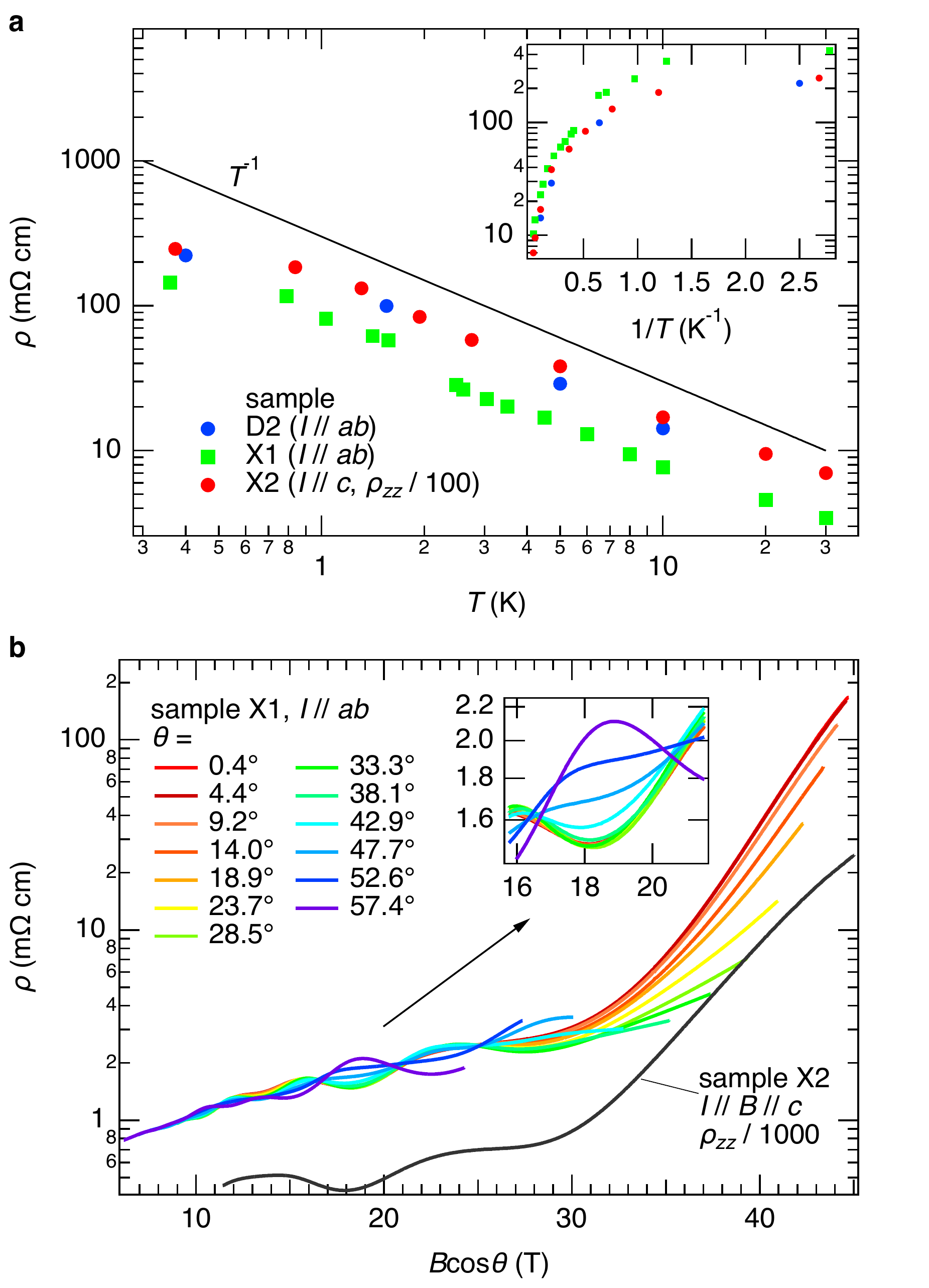}
\caption{\label{Fig4}
\textbf{Temperature and field-direction dependences of the resistivity in CaFeAsF.}  
\textbf{a} Resistivity at $B$ = 45 T as a function of temperature. The field direction is $B \parallel c$ for samples D2 and X2 and $\theta$ = 9.2 $^{\circ}$ for X1. The inset shows an Arrhenius plot of the same data. \textbf{b} Resistivity of sample X1 for various field angles plotted against the $c$-axis component of the applied field $B \cos \theta$. A blow-up of a region near $B = 19$ T is shown in the inset.  Note that the spin reduction factor $R_{\mathrm{s}}$ changes sign at $\theta \approx 50^{\circ}$ for both $\alpha$ and $\beta$ oscillations (see text for explanation).  The $c$-axis resistivity for $B \parallel c$ in sample X2 is also shown in the main panel.
}
\end{figure*}

\begin{table}[t]
\caption{\label{Tab} 
\textbf{Fermi-surface parameters.}
The first three parameters $F$, $m^*$, and $T_{\mathrm{D}}$ were estimated from the Shubnikov--de Haas oscillations for $B \parallel c$ in sample D2.
The rest except $S$ were derived from these first three: $n$ is the carrier density, $v_{\mathrm{F}}$ the Fermi velocity, $E_{\mathrm{F}}$ the Fermi energy, and $\Gamma = \pi k_{\mathrm{B}} T_{\mathrm{D}}$ the Landau-level broadening.
The value of $n$ for the $\alpha$ cylinder refers to the sum of the carriers of the two cylinders that occur in the Brillouin zone.
To derive $E_{\mathrm{F}}$, we used the linear dispersion relation ($E_{\mathrm{F}} = \hbar v_{\mathrm{F}} k_{\mathrm{F}}$) for the $\alpha$ cylinder and a quadratic one ($E_{\mathrm{F}} = \hbar^2 k_{\mathrm{F}}^2 / (2m^*)$) for $\beta$.
The spin-splitting parameter $S$ for $B \parallel c$ is based on \cite{Terashima18PRX}.
}
\begin{ruledtabular}
\begin{tabular}{ccc}
FS cylinder & $\alpha$ & $\beta$\\
\hline
carrier type & Dirac electron & Schr\"odinger hole\\
$F$ (T) & 19.01(1) & 47.49(2) \\
$m^*/m_{\mathrm{e}}$ & 0.41(2) & 0.90(5) \\
$T_{\mathrm{D}}$ (K) & 4.7(3) & 2.9(2) \\
$n$ (10$^{-3}$ per Fe) & 1.39 & 1.74 \\
$n$ (10$^{19}$ cm$^{-3}$) & 2.14 & 2.68\\
$v_{\mathrm{F}}$ (10$^4$ m s$^{-1}$) & 6.8 & 4.9\\
$E_{\mathrm{F}}$ (meV) & 10.7 & 6.1\\
$\Gamma$ (meV) & 1.3 & 0.78\\
$S$ & 0.321 & 1.61\\
\end{tabular}
\end{ruledtabular}
\end{table}

\end{document}